\documentclass[%
 reprint,
superscriptaddress,
 amsmath,amssymb,
 aps,
]{revtex4-2}

\usepackage{graphicx}
\usepackage{dcolumn}
\usepackage{bm}
\usepackage{amsmath}
\usepackage{lineno}
\usepackage[english]{babel}
\usepackage[utf8]{inputenc}
\usepackage{color}
\usepackage{siunitx}
\usepackage{amsmath}
\usepackage{textcomp, gensymb}
\usepackage{upgreek}
\usepackage{hyperref}
\hypersetup{
    colorlinks=true,
    citecolor=magenta,
    linkcolor=blue,
    filecolor=magenta,      
    urlcolor=magenta,
    }

\usepackage{subcaption}

\begin{document}

\title{Buried germanium quantum well proximitised by magnetic field-resilient superconducting platinum iridium germanosilicide}

\author{Praveen Viswanathan}
\affiliation{QuTech and Kavli Institute of Nanoscience, Delft University of Technology, Lorentzweg 1, 2628 CJ Delft, Netherlands}
\author{Erik Lemmens Sj\"ostrand}
\affiliation{QuTech and Kavli Institute of Nanoscience, Delft University of Technology, Lorentzweg 1, 2628 CJ Delft, Netherlands}
\author{Davide Costa}
\affiliation{QuTech and Kavli Institute of Nanoscience, Delft University of Technology, Lorentzweg 1, 2628 CJ Delft, Netherlands}
\author{Marinus Fischer}
\affiliation{Kavli Institute of Nanoscience, Delft University of Technology, Lorentzweg 1, 2628 CJ Delft, The Netherlands}
\author{Athique Ahmed}
\affiliation{Catalan Institute of Nanoscience and Nanotechnology – ICN2 (CSIC and BIST), Campus UAB, Bellaterra, Barcelona, Catalonia, Spain.}
\author{Sara Mart\'i-S\'anchez}
\affiliation{Catalan Institute of Nanoscience and Nanotechnology – ICN2 (CSIC and BIST), Campus UAB, Bellaterra, Barcelona, Catalonia, Spain.}
\author{Lucas E. A. Stehouwer}
\affiliation{QuTech and Kavli Institute of Nanoscience, Delft University of Technology, Lorentzweg 1, 2628 CJ Delft, Netherlands}
\author{Jordi Arbiol}
\affiliation{Catalan Institute of Nanoscience and Nanotechnology – ICN2 (CSIC and BIST), Campus UAB, Bellaterra, Barcelona, Catalonia, Spain.}
\affiliation{ICREA, Pg. Lluis Companys, Barcelona, Catalonia, Spain}
\author{Anasua Chatterjee}
\affiliation{QuTech and Kavli Institute of Nanoscience, Delft University of Technology, Lorentzweg 1, 2628 CJ Delft, Netherlands}
\author{Giordano Scappucci}
\affiliation{QuTech and Kavli Institute of Nanoscience, Delft University of Technology, Lorentzweg 1, 2628 CJ Delft, Netherlands}
\email{g.scappucci@tudelft.nl}
\author{Karina L. Hudson}
\email{k.l.hudson@tudelft.nl}
\affiliation{QuTech and Kavli Institute of Nanoscience, Delft University of Technology, Lorentzweg 1, 2628 CJ Delft, Netherlands}


\begin{abstract}
Hybrid superconductor-semiconductor systems provide a versatile platform for quantum technologies, ranging from superconducting-spin interfaces to topological quantum devices. Progress toward scalable implementations requires superconductors that exhibit high critical fields ($>\SI{1}{\tesla}$) at accessible temperatures integrated with low-disorder semiconductor heterostructures. Here we demonstrate a superconducting platinum iridium germanosilicide (PtIrSiGe), with critical out-of-plane magnetic field up to $B_{\perp} = \SI{1.9}{\tesla}$ and critical temperature of $T_c\sim \SI{1.85}{\kelvin}$, integrated with planar germanium with mobility $\mu = 1.3\times 10^6 \si{\centi\meter\squared\per\volt\per\second}$ via top-down lithography fabrication. We show that the integrity of the germanium quantum well and mobility and density of the 2D hole gas are preserved despite annealing at $\SI{500}{\celsius},$ a temperature comparable to that used for strained germanium epitaxy. We further demonstrate proximitisation of a buried germanium quantum well in a gate-defined Josephson junction/SQUID on a Ge/SiGe heterostructure. 
\end{abstract}

\maketitle

\vspace{10mm}
Planar germanium is emerging as a versatile platform to engineer a wide variety of quantum devices due to high hole mobility, low percolation density, tunable spin-orbit properties, and amenability to isotopic purification \cite{ScappucciNRM21}. It has become a frontrunning platform for quantum dot (QD) based spin qubits for engineering spin-based quantum processors\cite{WatzingerNatComms18,HendrickxNatComms20}, but has also found further application to quantum technology and condensed matter. The demonstration of proximitized superconductivity to a germanium quantum well via platinum germanosilicide (PtSiGe) \cite{TosatoNatComms23}, and further Andreev bound states in Josephson junctions\cite{KatePRA25} and in gate-controlled proximitization in a quantum dot\cite{Lakic2025Apr}, has broadened the potential applications of germanium toward hybrid semiconductor-superconductor devices\cite{LutchynNatRM18, FlensbergNatRM21, ChoiPRB00, LeijnsePRL13}. These include Kitaev chains\cite{vanLoo2026,tenHaaf2025}, gatemons\cite{CasparisNatNano18,sagi2024}, protected superconducting qubits including Andreev spin qubits (ASQs)\cite{Hays2021} and cos2$\phi$ qubits\cite{Larsen2020}, as well as fundamental experiments exploring the superconductor-insulator transition\cite{Bottcher2018} or in combining superconductivity and the quantum Hall effect\cite{Vignaud2023}, which have been predominantly explored in Group III-V materials or in graphene, where issues such as lower mobilities, less scalable fabrication, and coherence-limiting nuclear spins arise.

Previously, proximitisation of germanium to PtSiGe was achieved via top down lithography fabrication techniques compatible with existing semiconductor-based qubit device fabrication\cite{TosatoNatComms23}. While other methods have shown encouraging results, such as integrating an aluminium thin film or in granular form into the heterostructure via etching\cite{Aggarwal2021Apr} or using an ultra-shallow well\cite{Valentini2024Jan, FabrisAXV26}, PtSiGe integration does not require exposure of the well to air or introduction of crystal defects due to etching, and is compatible with deep, high-mobility, quantum wells\cite{Stehouwer2023Aug, Costa2024Nov, Costa2026Mar}. However, thin film PtSiGe has shown in experiments critical magnetic fields $B_{c,\perp} \sim \SI{100}{\milli\tesla}$ and $B_{c,\parallel} \sim \SI{400}{\milli\tesla}$\cite{TosatoNatComms23}, smaller than the magnetic fields typically required to resolve spin effects. The ability to use higher out-of-plane magnetic-fields with germanium quantum wells would also aid spin coherence due to the larger out-of-plane $g$-factor, enabling easier spin blockade readout\cite{Hendrickx2024}. Along with this, a larger superconducting parent and induced gap would alleviate quasiparticle poisoning effects, as well as enlarge the operational window for an array of protected qubits.

More robust superconductivity has been demonstrated using epitaxially grown tantalum germanide, with an out-of-plane critical field $B_{c,\perp} = \SI{1.88}{\tesla}$ and $T_c = 1.8 \sim\SI{2}{\kelvin}$ \cite{StrohbeenAPL24}; however, it is less easily integrated into lithographic fabrication methods. The aforementioned granular aluminium is also a robust superconductor compatible with lithographic fabrication, however, such proximitisation of a germanium quantum well requires an extremely thin Si$_{0.3}$Ge$_{0.7}$ spacer of $\SI{4}{\nano\meter}$ to ensure overlap of the superconducting wavefunction with the quantum well, and consequently the mobility of the 2D hole gas is impaired by surface defects. A recent systematic study of superconducting platinum group germanide and germanosilicide films reported critical fields of up to $B_{c,\perp} \sim \SI{2.8}{\tesla}$ and $T_c \sim \SI{2.6}{\kelvin}$\cite{LiAPL26} for IrSiGe, indicating its suitability for hybrid devices. However, this study did not demonstrate proximitisation of quantum wells, or integration into coherent nanoscale quantum devices.

In this work we demonstrate superconductivity of thin film platinum iridium germanosilicide (PtIrSiGe) at temperatures achievable in $^4$He cryostats, and an out-of-plane critical field approaching $\SI{2}{\tesla}$. We applied top-down lithography fabrication processes to fabricate a standard Hall bar and superconducting quantum interference device (SQUID) on a $\SI{25}{\nano\meter}$ deep germanium quantum well retaining mobility of order $10^6\si{\centi\meter\squared\per\volt\per\second}$. We provide high angle annular dark field scanning transmission electron microscopy (HAADF-STEM) analysis of the composition of the device stack, which shows that the buried germanium quantum well remains structurally pristine despite the high annealing temperature and complex thermal diffusion process resulting in PtIrSiGe forming uneven contact to the quantum well. Via low-temperature transport measurements, we extract density and mobility data from a Hall bar fabricated with PtIrSiGe contacts to verify that the integrity of the quantum well is preserved. Finally, we demonstrate proximitised superconductivity across the Josephson junctions in the SQUID, a key ingredient for superconducting-semiconducting hybrid qubits and beyond.

\section{Methods}

The Hall bar and SQUID devices were fabricated on an undoped accumulation mode Ge/Si$_{0.2}$Ge$_{0.8}$ quantum well heterostructure on a germanium substrate using standard electron beam lithography techniques. 
The two-dimensional hole gas (2DHG) forms in the compressively-strained Ge quantum well, positioned $\SI{25}{\nano\meter}$ below the dielectric interface. The PtIrSiGe contact to the quantum well is made by first removing the native oxide on the surface of the wafer using a buffered HF etch, then depositing $\SI{2}{\nano\meter}$ of Pt, followed by $\SI{8}{\nano\meter}$ of Ir via electron beam physical vapor deposition, and finally rapid thermal annealing at $\SI{500}{\celsius}$ for $5$ minutes in an argon atmosphere. Carrier accumulation is via a Ti/Pd top gate separated from the heterostructure by $\SI{20}{\nano\meter}$ of Al$_2$O$_3$ deposited via atomic layer deposition.

SQUID measurements were performed on a dilution refrigerator with a base temperature of $\SI{18}{\milli\kelvin}$ and a $\SI{1}{\tesla},\SI{1}{\tesla},\SI{6}{\tesla}$ vector magnet, where $\SI{6}{\tesla}$ is out-of-plane to the sample. Alongside the SQUID device on the same chip is a rectangular region of thin film used to verify the magnetic field and temperature response of the superconducting PtIrSiGe (thin film sample a). Additionally we show results from a separate test thin film of PtIrSiGe on Ge/SiGe quantum well heterostructure on a silicon substrate using the same fabrication process (thin film sample b) measured on a dilution refrigerator with a base temperature of $\SI{84}{\milli\kelvin}$ and a $^4$He refrigerator with a base temperature of $\SI{1.7}{\kelvin}$, both with a single-axis magnet.

All electrical measurements reported are performed with a four-terminal setup, using standard dc and low-frequency lock-in techniques.


\section{Thin film characterisation}

\begin{figure}
    \centering
        \includegraphics[width=\linewidth]{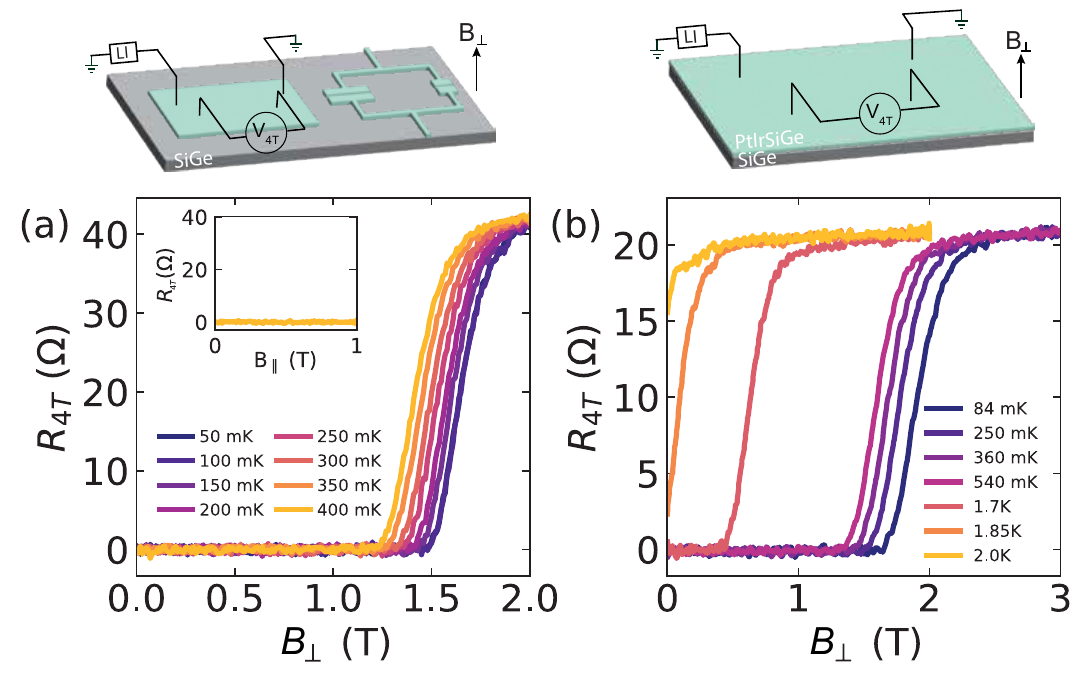} 
    \caption{(a) Thin film sample a: Resistance $R_{4T}$ vs out-of plane magnetic field $B_{\perp}$ for a range of temperatures. Inset: resistance $R_{4T}$ measured as a function for in-plane magnetic fields $B_{\parallel}$ up to $\SI{1}{\tesla}$ at $\SI{400}{\milli\kelvin}$. (b) Thin film sample b: $R_{4T}$ vs $B_{\perp}$ for varying temperature, measured in two separate cool downs in a dilution refrigerator and a $^4$He refrigerator.}
    \label{Fig1}
\end{figure}

To characterise the superconducting properties of the PtIrSiGe alloy, we studied the electrical transport of PtIrSiGe thin films.
Figure \ref{Fig1} shows the four-terminal resistance $R_{\mathrm{4T}}$ measured as a function of out-of-plane magnetic field $B_{\perp}$ for varying temperatures for thin film sample a and sample b in panels (a) and (b), respectively. These measurements were performed by applying a constant source ac current of $\SI{1}{\nano\ampere}$ at $\SI{7.777}{\hertz}$ and measuring the voltage drop between the source and drain contacts via four-terminal voltage probes as illustrated in the cartoons above each panel. Sample a shows superconductivity up to a critical field of $B_{\perp,c}=\SI{1.7}{\tesla}$ at $\SI{50}{\milli\kelvin}$, with the critical magnetic field decreasing with increasing sample temperature. The inset in this panel shows that the sample, at $\SI{400}{\milli\kelvin}$, remained superconducting up to an applied in-plane magnetic field $B_{\parallel}$ of $\SI{1}{\tesla}$, which was the maximum field available in our vector magnet set-up. Sample b shows superconductivity up to $B_{\perp,c}=\SI{1.9}{\tesla}$ at $\SI{84}{\milli\kelvin}$ and a critical temperature $T_c=\SI{1.85}{\kelvin}$, above the base temperature achievable in a $^4$He refrigerator. 
These robust superconducting properties can be attributed to the PtIrSiGe alloy -- bulk platinum has no measurable transition temperature\cite{RaubMatDes84}, and iridium has a critical temperature $\SI{112.5}{\milli\kelvin}$ \cite{GubserThermodynamic73}, well below that demonstrated in Fig. \ref{Fig1}. Moreover, in Ref. [~\onlinecite{Lakic2025Apr}], critical fields of close to $\SI{1}{\tesla}$ were observed in a PtSiGe nanoscale device designed to have leads of width $\sim$40nm, and it is conceivable that this method could further enhance the magnetic field resilience of PtIrSiGe.

\begin{figure*}
    \centering
        \includegraphics[width=\linewidth]{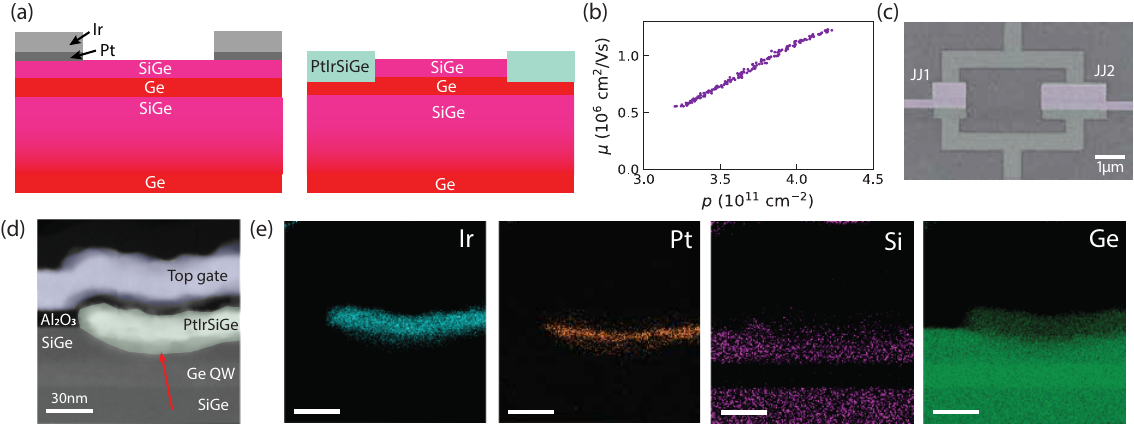} 
    \caption{(a) Schematic of the fabrication process. After removing the native oxide, platinum is deposited followed by iridium on the surface of the SiGe heterostructure. Rapid thermal annealing results in the Pt and Ir diffusing into the SiGe barrier to form PtIrSiGe alloy and contact to the germanium quantum well.(b) Mobility $\mu$ vs hole density $p$ of a Hall bar fabricated with PtIrSiGe contacts on the Ge/SiGe quantum well on germanium wafer. (c) False colour SEM of the SQUID device. The grey area is bare SiGe, the pale green regions are superconducting PtIrSiGe with Josephson junctions (JJs) 1 and 2 indicated. The pink regions are independent Ti/Pd top gates overlapping the superconductor-normal-superconductor Josephson junctions. (d) False colour cross-sectional HAADF-STEM image taken across the smaller Josephson junction in (c). The PtIrSiGe alloy thermally diffuses into the heterostructure, making contact with the quantum well at the position indicated by the red arrow. (e) Energy dispersive x-ray (EDX) spectroscopy images of the same cross-section in (d), showing the elemental mapping for iridium, platinum, silicon and germanium.}
    \label{Fig2}
\end{figure*}

\section{PtIrSiGe structural composition}

We now proceed to integrate these promising thin-film properties into the germanium quantum well, and additionally check that the quantum well remains undamaged. The process of forming contacts, done by annealing the sample at $\SI{500}{\celsius}$, is potentially degrading to the Ge/SiGe heterostructure, where the strained germanium quantum well itself is grown at the same temperature \cite{Stehouwer2023Aug}. Figure \ref{Fig2} (a) is a schematic of the cross-section of the Ge/SiGe heterostructure before and after annealing. We note that the calibration of the temperature readout of individual rapid thermal annealing tools is variable, and that the true annealing temperature in our system was likely below $\SI{500}{\celsius}$. Nevertheless the annealing temperature required for PtIrSiGe is much closer to the epitaxial growth temperature of the quantum well than that for PtSiGe. To verify that the high annealing temperature does not damage the integrity of the quantum well, we show in Figure \ref{Fig2} (b) the density and mobility characterisation on a standard Hall bar structure, with PtIrSiGe contacts, fabricated in parallel with the SQUID device. We obtain a maximum mobility of $\mu = 1.3\times 10^6 \si{\centi\meter\squared\per\volt\per\second}$ at density $p = 4.3\times10^{11}\si{\centi\meter\squared}$. For comparison, Hall bar devices fabricated with normal ohmic contacts on a $\SI{55}{\nano\meter}$ deep germanium quantum well grown on germanium substrate report hole mobilities of up to $\mu = 3\times 10^6 \si{\centi\meter\squared\per\volt\per\second}$\cite{Stehouwer2023Aug}, and 
on a $\SI{22}{\nano\meter}$ deep germanium quantum well grown on silicon substrate where strain defects are more prevalent up to $\mu = 5\times 10^5 \si{\centi\meter\squared\per\volt\per\second}$\cite{SammakAFM19}.

Figure \ref{Fig2} (c) is a false colour scanning electron microscope image of the measured SQUID. The grey region is bare wafer, pale green region is PtIrSiGe, and the violet region indicates the top gate. Junction 1 is $\SI{1}{\micro\meter}$ wide, and junction 2 is $\SI{2}{\micro\meter}$ wide, and both junctions are $\SI{70}{\nano\meter}$ long. HAADF-STEM analysis was performed in a cross-section taken across junction 1 of the SQUID to examine the structure and is shown in Fig. \ref{Fig2} (d), revealing that the Pt/Ir has diffused unevenly into the SiGe, making contact with the Ge quantum well in the region indicated by the red arrow, and in other regions remaining separated from the quantum well by the SiGe barrier. Fig. \ref{Fig2} (e) contains energy dispersive x-ray (EDX) spectroscopy data showing the distribution of iridium, platinum, silicon, and germanium in the cross-sectional area in (d). The platinum and iridium have diffused downwards in the region indicated by the red arrow in (d), while the germanium and some silicon have diffused upwards to form the PtIrSiGe layer. See the Supplementary Material for additional electrical characterisation and HAADF-STEM data of PtIrSiGe.

The choice to include a layer of $\SI{2}{\nano\meter}$ of platinum between the SiGe heterostructure and the $\SI{8}{\nano\meter}$ of iridium was motivated by the uneven diffusivity of germanium and iridium which were observed in our early trials of thin film IrSiGe. These samples exhibit $B_{c,\perp}$ up to $\sim\SI{1.2}{\tesla}$ at $T=\SI{1.7}{\kelvin}$. However, HAADF-STEM analysis of the IrSiGe sample revealed severe structural damage to the SiGe heterostructure. There was minimal diffusion of the iridium into the SiGe heterostructure. Instead, germanium and some silicon were drawn out of the heterostructure to form a layer of IrSiGe on the surface of the wafer, resulting in the formation of voids at the interface of IrSiGe and the SiGe heterostructure. Our observation of outward diffusion of germanium and silicon into the iridium (in contrast to inward diffusion of platinum into SiGe\cite{TosatoNatComms23}) is consistent with a previous study showing platinum silicide formation is driven by inward platinum diffusion and iridium silicide by outward silicon migration\cite{MorganASS92}. The inclusion of platinum was therefore designed to mediate the diffusion process and prevent the formation of voids \cite{Rahman2005Oct}. The HAADF-STEM images in Fig. \ref{Fig2} (d) and (e) do not show voids forming at the semiconductor-superconductor interface, and importantly the PtIrSiGe alloy diffuses downwards in some regions to contact the germanium quantum well. The pitted surface morphology of the PtIrSiGe is also visible under atomic force microscopy (AFM), while the surface of IrSiGe remains comparatively even (see Supplementary Material).
We note that the uneven diffusion of PtIrSiGe in Fig. \ref{Fig2} (d) appears qualitatively similar in form to the cross-section HAADF-STEM of IrGe encapsulated in platinum in Ref. [\onlinecite{LiAPL26}].

\section{Proximitising germanium to PtIrSiGe}

\begin{figure}
    \centering
        \includegraphics[width=\linewidth]{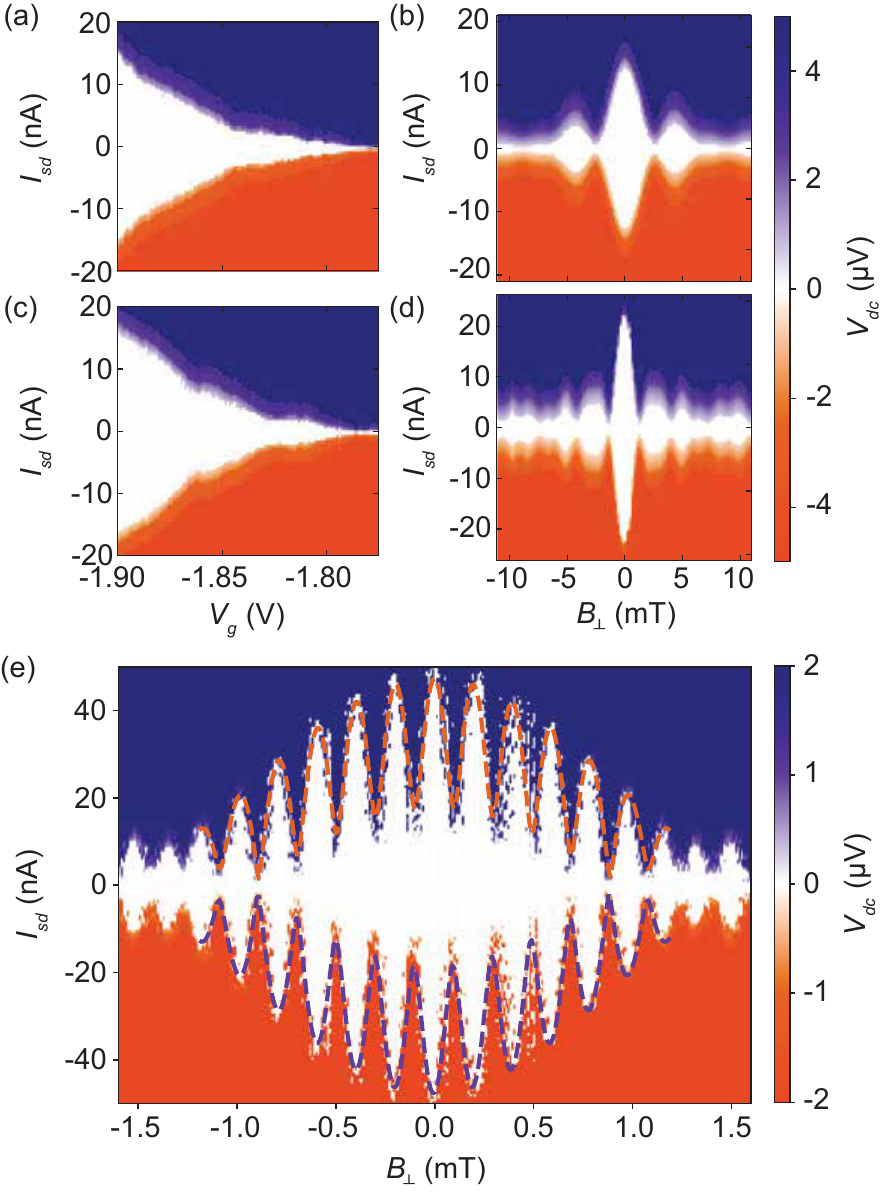} 
    \caption{
    \textbf{Junction 1:} (a) Colour map of dc voltage drop $V_{dc}$ across Josephson junction 1 as a function of dc source-drain current $I_{sd}$ and top gate voltage $V_g$. (b) Colour map of $V_{dc}$ as a function of dc $I_{sd}$ and out-of-plane magnetic field $B_{\perp}$ at fixed $V_g = \SI{-1.9}{\volt}$. 
    \textbf{Junction 2:} (c) Colour map of $V_{dc}$ across Josephson junction 2 as a  function of dc $I_{sd}$ and $V_g$. (d) Colour map of $V_{dc}$ as a function of $I_{sd}$ and $B_{\perp}$ at fixed $V_g = \SI{-1.9}{\milli\volt}$.
    (e) SQUID oscillations with both junctions conducting, such that $I_{c2}\approx 2 I_{c1}$. Fits to equation (1) are shown as dashed lines, in orange for the positive branch and purple for the negative branch. (b), (d), and (e) are offset in $B_{\perp}$ to account for known hysteresis in the magnet.
    }
    \label{Fig3}
\end{figure}

The electrical properties of the SQUID (Fig.~\ref{Fig2}(c)) are shown in Figure \ref{Fig3}. We are able to control the switching current across each junction via independent top gates. Fig.~\ref{Fig3}(a) is a colormap of four-terminal voltage $V$ measured across junction 1 as a function of source-drain dc current $I_{dc}$ and top gate voltage $V_{g,1}$, with $V_{g,2} = \SI{0}{\volt}$. Note that at zero gate voltage, as expected in an accumulation-mode heterostructure, there is no transport across junction 2. The purple region corresponds to positive voltage and orange to negative voltage, and white to $V=\SI{0}{\volt}$ drop across the junction. As $V_{g,1}$ is tuned to more negative values, corresponding to a higher density of holes and more transport channels in the Ge quantum well, the switching current in $I_{dc}$ increases. Fig.~\ref{Fig3}(b) is a colour map of $V$ measured across junction 1 as a function of $I_{dc}$ and applied out-of-plane magnetic field $B_{\perp}$ for $V_{g,1} = \SI{-1.9}{\volt}$. The white superconducting region exhibits Fraunhofer-like oscillations centered and symmetric around $B_{\perp} = \SI{0}{\tesla}$ (corrected for a known hysteresis offset in the magnet). The oscillations do not exhibit the characteristic single-period Fraunhofer pattern, which we attribute to the variable diffusion depth of the PtIrSiGe giving rise to an effective Josephson junction geometry in which the length varies along its width. Fig. \ref{Fig3} (c) and (d) show similar gate-tunable switching current and Fraunhofer-like oscillations for junction 2, for which junction 1 was set to $V_{g,1} = \SI{0}{\volt}$. Junction 2 being approximately twice as wide, has a smaller oscillation period in $B_{\perp}$ compared to junction 1 (see Supplementary Material Section III for Fraunhofer fits).

We investigate the SQUID operation by setting the junctions in an asymmetric regime, such that the critical current through junction 2 is twice as large as that through junction 1, i.e., $I_{c2} \approx 2 I_{c1}$. We then observe SQUID oscillations as smaller oscillations on top of the Fraunhofer envelope of the larger junction, modulated by the applied out-of-plane magnetic field. We fit these oscillations to the relation \cite{Fominov2022Oct}
\begin{equation} \label{eq:fraunhofer}
\begin{split}
        I_c = \sqrt{
        \left( I_{c1} - I_{c2} \right) ^2
        + 4 I_{c1} I_{c2} \cos^2 \left( \frac{\pi B_\perp A_{SQUID}}{\Phi_0} \right)}
\end{split}
\end{equation}
where $I_{c1,2} (B A_{1,2})$ are the Fraunhofer dependencies of the critical current obtained from fitting the Fraunhofer pattern of each junction, $A_{1,2}$ are the individual junction areas, $B_{\perp}$ is the out-of-plane magnetic field, and $\Phi_0$ is the flux quantum.
Note that due to the low critical currents and even lower circulating current, the phase contribution due to self-inductance was negligible compared to that  due to the external magnetic field, and the self-inductance could not be extracted from the experimental data.
From the fit of the positive (negative) branch in Fig. \ref{Fig3} (e) to equation \ref{eq:fraunhofer}, we extract the effective SQUID loop area $A_{SQUID} = \SI{10.47}{\micro\meter\squared}$ ($\SI{10.48}{\micro\meter\squared}$), which is in reasonable agreement with the geometric area $\SI{10}{\micro\meter\squared}$, the slight increase likely arising due to flux-focussing effects \cite{GranataPR2016}.

\section{Discussion}

We have demonstrated superconductivity in thin film PtIrSiGe with a robust out-of-plane critical field approaching $\SI{2}{\tesla}$ and critical temperature of $\sim\SI{1.85}{\kelvin}$. We note the critical field and temperature of PtIrSiGe is comparable to reported values of superconducting thin film IrGe and IrSiGe \cite{LiAPL26}, and that we also observe a similar breakdown of superconductivity if we anneal the PtIrSiGe at a sufficiently high temperature of $\SI{550}{\celsius}$ (see Supplementary Material). The performance of thin film PtIrSiGe is also comparable to epitaxially grown tantalum germanide\cite{StrohbeenAPL24}, and granular aluminium on ultra-shallow germanium quantum well heterostructure\cite{FabrisAXV26}, where the critical out-of-plane fields are also $>\SI{1}{\tesla}$, with critical temperatures close to $\SI{2}{\kelvin}$. Further process development is required to establish a uniform and reproducible interface between the PtIrSiGe and the quantum well. We speculate that increasing the ratio of platinum to iridium may result in a superconductor with a reduced critical field but more reliable and uniform PtIrSiGe layer. 

We have demonstrated proximitisation of a buried germanium quantum well to superconducting PtIrSiGe without compromising the mobility of the 2D hole gas and using simple fabrication methods. The platinum and iridium layers are deposited successively in the same evaporator, and we do not employ additional etching steps or fabrication tools beyond our standard process for ohmic-contacted devices. Aside from our choice to use wafer with a relatively shallow $\SI{25}{\nano\meter}$ deep germanium quantum well, our SiGe heterostructure is otherwise similar to that used for non-hybrid quantum devices. Crucially, further work is needed to establish the size and quality of the superconducting gap. 

While these results represent a proof of principle demonstration of PtIrSiGe in hybrid semiconductor-superconductor devices, they also mark a technological milestone by demonstrating superconducting proximitisation of an ultra-low-disorder, high-mobility germanium quantum well to a superconducting layer with large critical fields and accessible critical temperature, patterned using top-down lithographic fabrication techniques.

\bigskip

\small{We acknowledge support by the European Union through the IGNITE project with grant agreement No. 101069515 and the QLSI project with grant agreement No. 951852. A.C. and P.V. acknowledge support from the European Union through the ELEQUANT project (grant agreement No. 101185712) and the TU Delft Excellence Fund.
This work was supported by the Netherlands Organisation for Scientific Research (NWO/OCW), via the Open Competition Domain Science - M program. We acknowledges the research programme Materials for the Quantum Age (QuMat) for financial support. This programme (registration no. 024.005.006) is part of the Gravitation programme financed by the Dutch Ministry of Education, Culture and Science (OCW). This research was sponsored in part by The Netherlands Ministry of Defence under Awards No. QuBits R23/009. The views, conclusions, and recommendations contained in this document are those of the authors and are not necessarily endorsed nor should they be interpreted as representing the official policies, either expressed or implied, of The Netherlands Ministry of Defence. The Netherlands Ministry of Defence is authorized to reproduce and distribute reprints for Government purposes notwithstanding any copyright notation herein. This research was supported in part by the Army Research Office (Grant No. W911NF-17-1-0274). 
The views and conclusions contained in this document are those of the authors and should not be interpreted as representing the official policies, either expressed or implied, of the Army Research Office (ARO), or the U.S. Government. 
The U.S. Government is authorized to reproduce and distribute reprints for Government purposes, notwithstanding any copyright notation herein.

J.A. acknowledges ICN2 and the Joint Electron Microscopy Center at ALBA (JEMCA) for providing key (S)TEM and FIB facilities and technical guidance. ICN2 is supported by the Severo Ochoa program from Spanish MCIN/AEI (Grant No. CEX2021-001214-S), and the CERCA Programme, Generalitat de Catalunya. ICN2 is founding member of e-DREAM. ICN2 acknowledges funding from Generalitat de Catalunya (Grant No. 2021SGR00457). We acknowledge support from CSIC Interdisciplinary Thematic Platform (PTI+) on Quantum Technologies (PTI-QTEP+). This work has been funded by the European Commission – NextGenerationEU (Regulation EU 2020/2094), through CSIC's Quantum Technologies Platform (QTEP). ICN2 acknowledges funding from Grant IU16-014206 (METCAM-FIB) funded by the European Union through the European Regional Development Fund (ERDF), with the support of the Ministry of Research and Universities, Generalitat de Catalunya. This work is in the framework of the Universitat Autonoma de Barcelona Materials Science PhD program.}

\subsection*{Authors contributions and declarations}
L.E.A.S. grew the Ge/SiGe heterostructure developed with input from G.S. A.A., S.M.S and J.A. performed analysis of the material stack. K.L.H. fabricated the devices with input from M.F. P.V., K.L.H., and E.L.S. performed measurements with input from D.C. P.V. and K.L.H. wrote the manuscript with input from A.C and G.S.

G.S. is founding advisor of Groove Quantum BV and declares equity interests.

\subsection*{Data availability statement}

The data sets supporting the findings of this study are
openly available at the Zenodo repository~Ref.~[\onlinecite{Viswanathan2026Aug}].

\bibliography{References}

\end{document}


\title{Buried germanium quantum well proximitised by magnetic field-resilient superconducting platinum iridium germanosilicide \\ Supplementary Material}

\author{Praveen Viswanathan}
\affiliation{QuTech and Kavli Institute of Nanoscience, Delft University of Technology, Lorentzweg 1, 2628 CJ Delft, Netherlands}
\author{Erik Lemmens Sj\"ostrand}
\affiliation{QuTech and Kavli Institute of Nanoscience, Delft University of Technology, Lorentzweg 1, 2628 CJ Delft, Netherlands}
\author{Davide Costa}
\affiliation{QuTech and Kavli Institute of Nanoscience, Delft University of Technology, Lorentzweg 1, 2628 CJ Delft, Netherlands}
\author{Marinus Fischer}
\affiliation{Kavli Institute of Nanoscience, Delft University of Technology, Lorentzweg 1, 2628 CJ Delft, The Netherlands}
\author{Athique Ahmed}
\affiliation{Catalan Institute of Nanoscience and Nanotechnology – ICN2 (CSIC and BIST), Campus UAB, Bellaterra, Barcelona, Catalonia, Spain.}
\author{Sara Mart\'i-S\'anchez}
\affiliation{Catalan Institute of Nanoscience and Nanotechnology – ICN2 (CSIC and BIST), Campus UAB, Bellaterra, Barcelona, Catalonia, Spain.}
\author{Jordi Arbiol}
\affiliation{ICREA Pg. Lluis Companys, Barcelona, Catalonia, Spain}
\affiliation{Catalan Institute of Nanoscience and Nanotechnology – ICN2 (CSIC and BIST), Campus UAB, Bellaterra, Barcelona, Catalonia, Spain.}
\author{Lucas E. A. Stehouwer}
\affiliation{QuTech and Kavli Institute of Nanoscience, Delft University of Technology, Lorentzweg 1, 2628 CJ Delft, Netherlands}
\author{Anasua Chatterjee}
\affiliation{QuTech and Kavli Institute of Nanoscience, Delft University of Technology, Lorentzweg 1, 2628 CJ Delft, Netherlands}
\author{Giordano Scappucci}
\affiliation{QuTech and Kavli Institute of Nanoscience, Delft University of Technology, Lorentzweg 1, 2628 CJ Delft, Netherlands}
\email{g.scappucci@tudelft.nl}
\author{Karina L. Hudson}
\email{k.l.hudson@tudelft.nl}
\affiliation{QuTech and Kavli Institute of Nanoscience, Delft University of Technology, Lorentzweg 1, 2628 CJ Delft, Netherlands}

\date{\today}


\maketitle

\newpage
\section{Additional characterisation of \NoCaseChange{PtIrSiGe}}

The superconductivity of PtIrSiGe thin film was tested for a range of annealing temperatures and times that were based on earlier trials of IrSiGe (which are outlined below in Supplementary Material Section II). The PtIrSiGe thin film shown in Figure 1 (b) of the main text was part of a larger piece of wafer on which Pt/Ir was deposited before being broken into several identical pieces that were then annealed at different temperatures and times. Fig. \ref{SFig1} is a plot of normalised resistance $\rho_N$ against out-of-plane magnetic field $B_{\perp}$ measured in a $^4$He refrigerator at $\SI{1.7}{\kelvin}$ for three such samples. We found that annealing at $\SI{500}{\celsius}$ for $\SI{5}{\minute}$ yielded superconductivity with an out-of-plane critical field $B_{\perp} \sim \SI{2}{\tesla}$ when measured at $T = \SI{80}{\milli\kelvin}$. Annealing for longer very marginally improved the critical field, and annealing at $\SI{550}{\celsius}$ likely partly induced a semiconducting crystal phase of PtIrSiGe which damaged the superconducting field range. We chose to anneal our Hall bar and SQUID devices at $\SI{500}{\celsius}$ for $\SI{5}{\minute}$ as this achieved good superconducting characteristics with minimal exposure to temperatures at the upper limit of the thermal budget of the Ge/SiGe heterostructure. 

\vspace{10mm}

{
\centering
    \includegraphics[width=\linewidth]{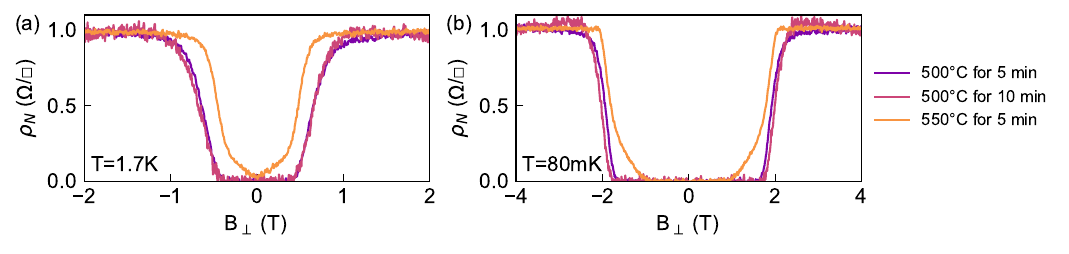} 
    \captionsetup{justification=raggedright,singlelinecheck=false}
    \captionof{figure}{Normalised resistivity $\rho_N$ of PtIrSiGe thin film vs out-of plane magnetic field $B_{\perp}$ for a range of annealing temperatures and times, measured at (a) $\SI{1.7}{\kelvin}$, and (b) $\SI{80}{\milli\kelvin}$.}
    \label{SFig1}
}
\vspace{5mm}

\begin{figure}
    \centering
        \includegraphics[width=\linewidth]{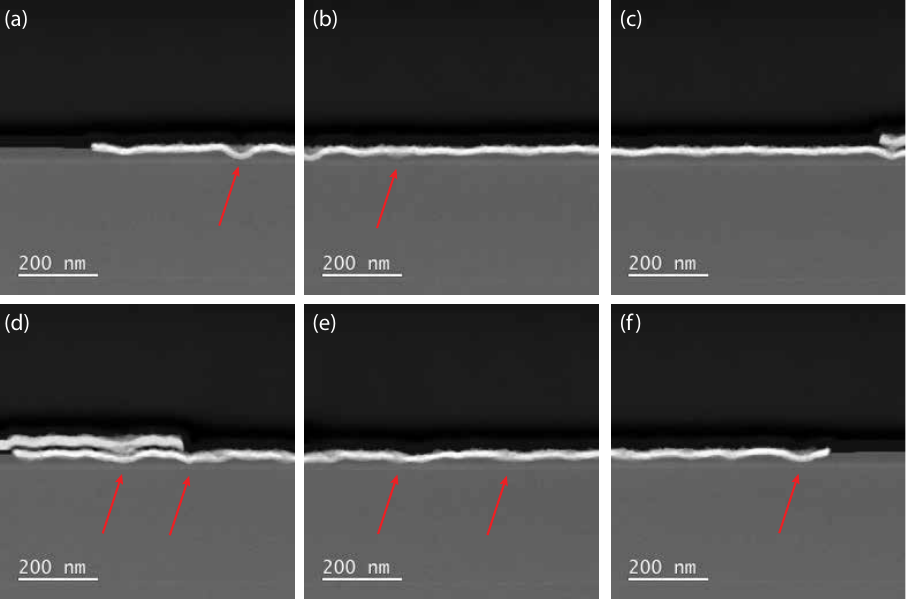} 
    \caption{STEM images showing the diffusion of PtIrSiGe along the length of the cross-section taken for analysis. The red arrows indicate the regions where the PtIrSiGe is touching the germanium quantum well.}
    \label{SFig2}
\end{figure}

In Figure \ref{SFig2} are six additional cross-sectional high-angle annular dark field scanning transmission electron microscope (STEM) images taken along with that shown in Fig. 2 (d) in the main text. Here the uneven downward diffusion of PtIrSiGe into the heterostructure is visibly apparent. The PtIrSiGe appears to make contact with the germanium quantum well only in the regions indicated by the red arrows.

\begin{figure}
    \centering
        \includegraphics[width=\linewidth]{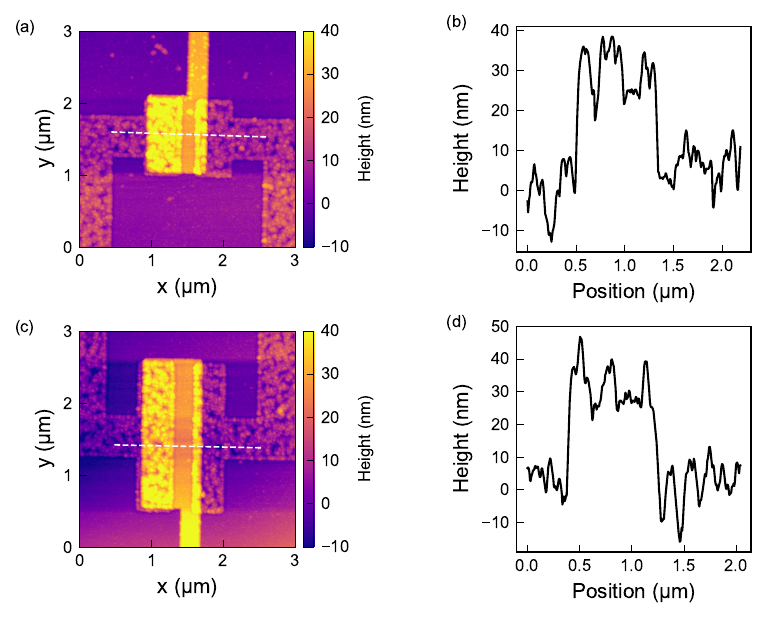} 
    \caption{(a) AFM image of Josephson junction 1. The dashed line indicates the region of cross section which is plotted in panel (b). (c) AFM image of Josephson junction 2, with the dashed line indicating the region of cross section displayed in panel (d).}
    \label{SFig3}
\end{figure}

Figure \ref{SFig3} shows atomic force microscopy (AFM) measurements of each Josephson junction of the SQUID described in the main text. Panel (a) is a colourmap of height with respect to $x$ and $y$ position. The pitted form of the PtIrSiGe layer is visible. A cross-section indicated by the white dashed line is plotted in panel (b). The AFM map was levelled so that the surface of the wafer is at height $0$. The PtIrSiGe level fluctuates up to $\pm\SI{10}{\nano\meter}$ above and below the surface of the SiGe wafer, consistent with the STEM images in Fig. 2(d) in the main text and Fig. \ref{SFig2} that shows some regions of PtIrSiGe diffuse downward into the SiGe and other regions remain on the surface of the SiGe heterostructure.

\FloatBarrier

\section{\NoCaseChange{IrSiGe} thin film}

Here we present electrical characterisation and some TEM results from an early trial of IrSiGe thin film for reference against PtIrSiGe in the main text.

\begin{figure}
    \centering
        \includegraphics[width=\linewidth]{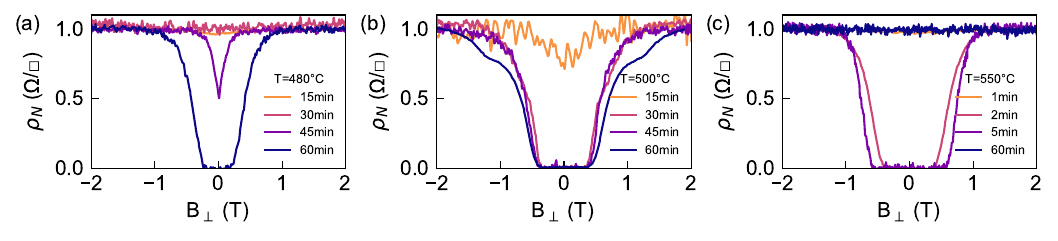} 
    \caption{Resistance $R$ vs out-of plane magnetic field $B_{\perp}$ up to $\SI{3}{\tesla}$ measured at $\SI{1.7}{\kelvin}$ for a range of annealing temperatures and times. }
    \label{SFig4}
\end{figure}

$\SI{10}{\nano\meter}$ of iridium was deposited on undoped accumulation mode Ge/Si$_{0.2}$Ge$_{0.8}$ quantum well heterostructure on a silicon wafer after removal of native oxides using standard optical lithography techniques. The two-dimensional hole gas (2DHG) forms in the compressively-strained Ge quantum well, positioned $\SI{20}{\nano\meter}$ below the surface. IrSiGe alloy was formed via rapid thermal annealing at a range of temperatures and times in an argon atmosphere in the same furnace used for the devices in the main text. Measurements were performed on a $^4$He refrigerator with a base temperature of $\SI{1.7}{\kelvin}$. The results are shown in Figure \ref{SFig4}: panel (a) shows normalised resistance $\rho_N$ against out-of-plane field $B_{\perp}$ measured at $\SI{1.7}{\kelvin}$ for IrSiGe thin films annealed at $\SI{480}{\celsius}$ for times ranging from $\SI{15}{\minute}$ to $\SI{60}{\minute}$. While superconductivity ($\rho_N=0$ at small magnetic field) can be achieved by annealing at $\SI{480}{\celsius}$, we chose to later investigate higher annealing temperatures to reduce the processing time. Panel (b) shows $\rho_N$ against $B_{\perp}$ for IrSiGe thin films annealed at $\SI{500}{\celsius}$ for times ranging from $\SI{15}{\minute}$ to $\SI{60}{\minute}$, where similar superconducting properties are observed as panel (a) for $\SI{30}{\minute}$ and $\SI{45}{\minute}$, and is also consistent with the results of Ref. [\onlinecite{LiAPL26}]. After $\SI{60}{\minute}$ the profile of the transition from superconducting to normal in magnetic field deviates from the shape observed for shorter annealing times, which we speculate is due to the formation of more than one crystal phase. Finally, in panel (c) we trialled annealing beyond the expected thermal budget of Ge/SiGe heterostructures at $\SI{550}{\celsius}$. We observe that no superconductivity is observed for $\SI{1}{\minute}$ anneal time, however $2$ and $\SI{5}{\minute}$ yield similar superconducting behaviour in magnetic field. However after $\SI{60}{\minute}$ the IrSiGe appears to have returned to a normal semiconducting crystal phase. 

Cross-sectional transmission electron microscope (TEM) of IrSiGe on Ge/SiGe after annealing at $\SI{500}{\celsius}$ is shown in Supplementary Figure \ref{SFig5} (a). The IrSiGe layer has not diffused into the heterostructure, and the SiGe spacer between the germanium quantum well and the IrSiGe layer remains largely intact. The irregularly shaped dark regions underneath the IrSiGe are voids that have formed due to germanium and some silicon diffusing upwards into the IrSiGe layer. The iridium has not diffused downwards towards the quantum well. This is shown clearly in the energy dispersive x-ray (EDX) spectroscopy data in panels (b)-(e), showing the position of iridium, germanium, silicon and oxygen in the stack in panel (a). The voids are mostly missing germanium, and some trace oxygen has diffused downwards to line the voids.

\begin{figure*}
    \centering
        \includegraphics[width=0.4\linewidth]{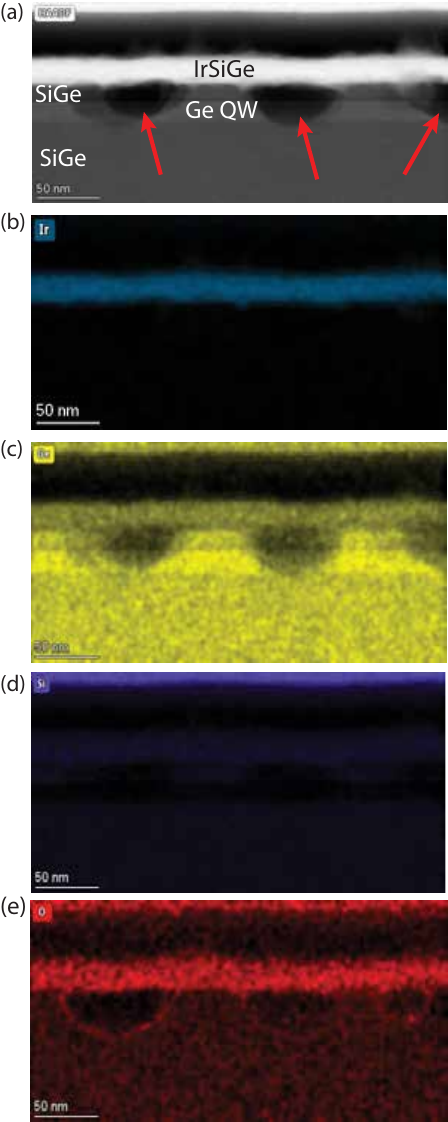} 
    \caption{(a) Cross-sectional transmission electron microscope image taken across a region of annealed IrSiGe on SiGe heterostructure. The  dark line between the Ge quantum well and the upper SiGe buffer is a $\SI{2}{\nano\meter}$ thick silicon spike. The dark voids indicated by red arrows form underneath the IrSiGe layer. (b) Energy dispersive x-ray spectroscopy images of the same cross-section in (a), showing the elemental mapping for iridium, (c) germanium, (d) silicon, and (e) oxygen.}
    \label{SFig5}
\end{figure*}

\begin{figure}
    \centering
        \includegraphics[width=\linewidth]{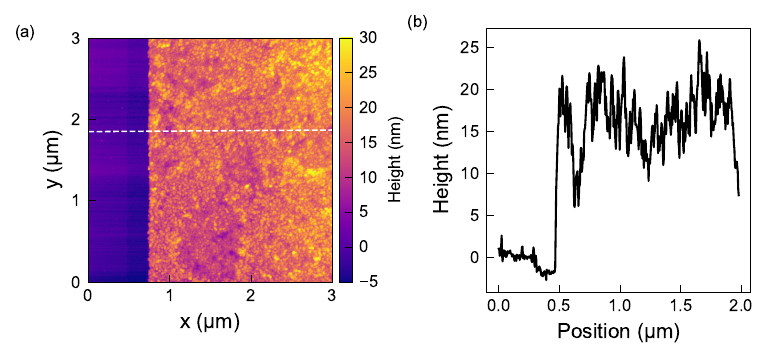} 
    \caption{(a) AFM image of a test IrSiGe deposition. On the left side is bare SiGe wafer, and the granular textured region is IrSiGe. The dashed line indicates the region of cross section which is plotted in panel (b). }
    \label{SFig6}
\end{figure}

Figure \ref{SFig6} shows atomic force microscopy (AFM) measurements of annealed IrSiGe. Panel (a) is a colourmap of height with respect to $x$ and $y$ position. Annealed IrSiGe has a granular surface morphology in contrast to the smooth surface of the SiGe heterostructure. A cross-section indicated by the white dashed line is plotted in panel (b). The AFM map was levelled so that the surface of the wafer is at height $0$. The IrSiGe level fluctuates but unlike PtIrSiGe in Fig. \ref{SFig4} does not dip below the surface level of the SiGe wafer and does not make physical contact with the germanium quantum well.

\FloatBarrier

\section{Fraunhofer fitting to Josephson junctions}

Here we show the fitting of the Fraunhofer-like oscillations of each Josephson junction from Figure 3 (b) and (d) of the main text. These fits are used to estimate the area of individual junctions, which are then used to fit the SQUID oscillations in Figure 3 (e).

\begin{figure}[!h]
    \centering
    \begin{subfigure}{0.48\textwidth}
        \includegraphics[width=\linewidth]{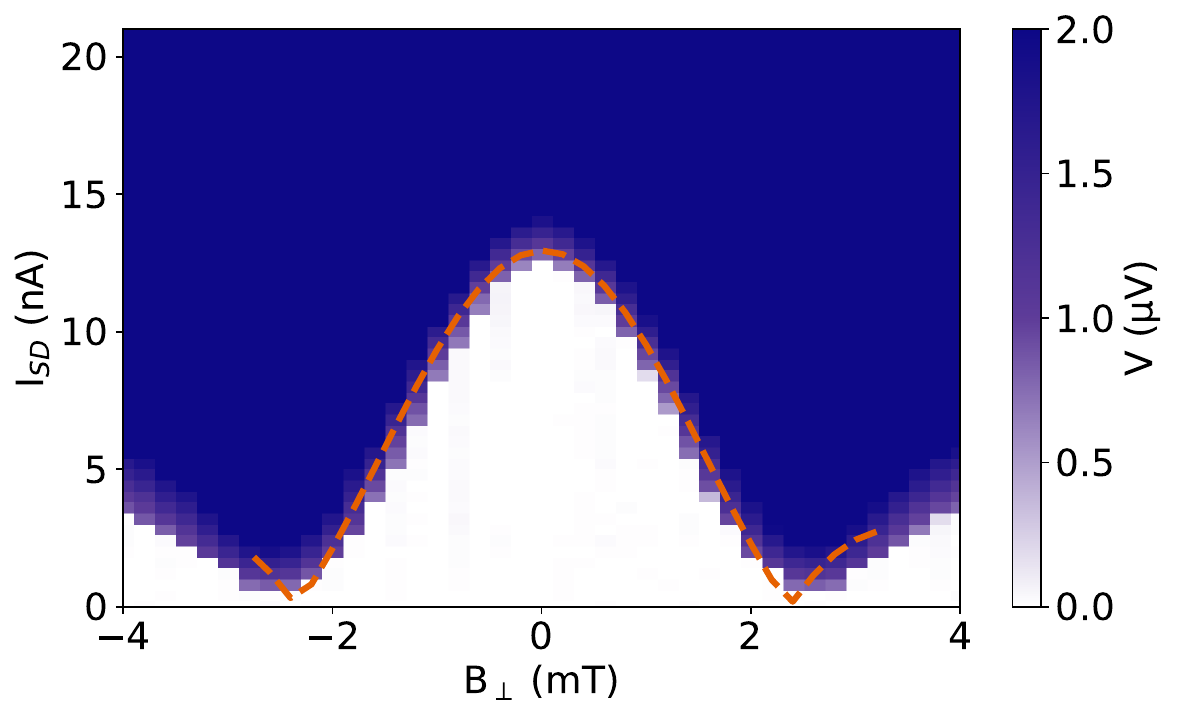}
        \caption{}
    \end{subfigure}
    \begin{subfigure}{0.48\textwidth}
        \includegraphics[width=\linewidth]{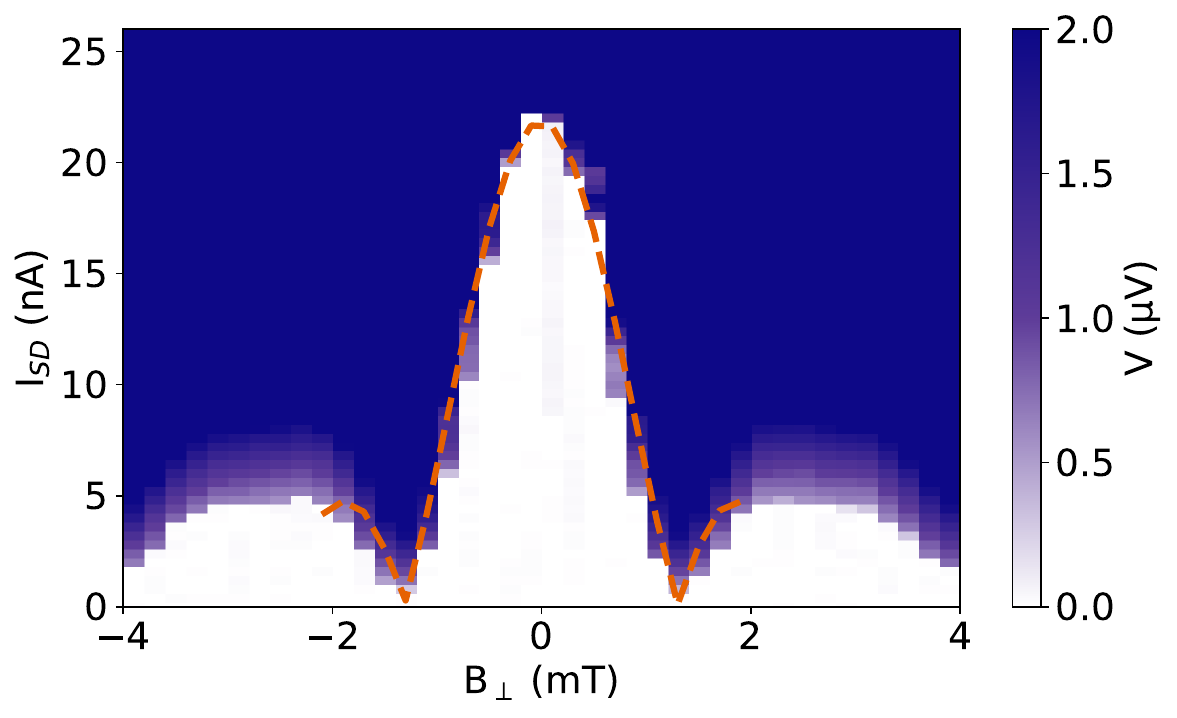}
        \caption{}
    \end{subfigure}
    \caption{
    Fraunhofer-like modulation of the switching current with applied external magnetic field $B_\perp$, and corresponding fit (dashed orange line) to equation \ref{eqn:fraunhofer} for (a) Junction 1, and (b) Junction 2.
    Note that the data here is cropped from Fig. 3 in the main text.
    }
    \label{SFig7}
\end{figure}

The expression for the critical currents $I_{c1,2}$ due to Fraunhofer interference oscillations is 
\begin{equation}
    I_{c1,2}(B_\perp A_{1,2}) =I_{c0\,1,2}\,\frac{\sin\left(\pi B_\perp A_{1,2}/\Phi_0\right)}{\pi B_\perp A_{1,2}/\Phi_0}
    \label{eqn:fraunhofer}
\end{equation}
where $B_\perp$ is the applied out-of-plane magnetic field, $A_{1,2}$ are the individual junction areas, and $\Phi_0$ is the flux quantum.
The uneven proximitisation of PtIrSiGe to the germanium quantum well presents a challenge in finding a reasonable estimation of the fitting parameters.
The fit presented in Fig. 
\ref{SFig7}, estimating junction areas $A_{1} = \SI{0.88}{\micro\meter\squared}$ and $A_{2} = \SI{1.58}{\micro\meter\squared}$ is our best estimation which we applied to characterise the SQUID, presented in Fig. 3 in the main text.

\FloatBarrier

\bibliography{References.bib}